\documentclass[11pt]{article} 

\usepackage[utf8]{inputenc} 
\usepackage{amsmath,amssymb,hyperref,graphicx,authblk,color,xcolor}
\usepackage{url,comment}

\newcommand{\intd}[1]{\int\!\!\diff #1}

\newcommand{\rr}{\mathbf{r}}

\newcommand{\uu}{\mathbf{u}}
\newcommand{\diff}{\mathrm{d}}

\newcommand{\jw}{\mathbf{j}_w}
\newcommand{\jh}{\mathbf{j}_p}
\newcommand{\Xh}{\mathbf{X}_p}
\newcommand{\Xw}{\mathbf{X}_w}
\newcommand{\Lhh}{L_{pp}}
\newcommand{\Lhw}{L_{pw}}
\newcommand{\Lwh}{L_{wp}}
\newcommand{\Lww}{L_{ww}}
\newcommand{\XX}{\mathbf{X}}
\newcommand{\JJ}{\mathbf{J}}
\newcommand{\XE}{{x^*}}
\newcommand{\YE}{{y^*}}
\newcommand{\ww}{\mathbf{w}}

\begin{document}
\title{Braun-Le Chatelier principle in dissipative thermodynamics}
\author[1]{Michal Pavelka\thanks{pavelka@karlin.mff.cuni.cz}}
\author[2]{Miroslav Grmela}
\affil[1]{Mathematical Institute, Faculty of Mathematics and Physics, Charles University in Prague, Sokolovsk\'{a} 83, 186 75 Prague, Czech Republic}
\affil[2]{\'{E}cole Polytechnique de Montr\'{e}al, C.P.6079 suc. Centre-ville, Montréal, H3C 3A7, Québec, Canada}

\maketitle
\begin{abstract}
Braun-Le Chatelier principle is a fundamental result of equilibrium thermodynamics, showing how stable equilibrium states shift when external conditions are varied. The principle follows from convexity of thermodynamic potential. Analogously, from convexity of dissipation potential it follows how steady non-equilibrium states shift when thermodynamic forces are varied, which is the extension of the principle to dissipative thermodynamics.\\
\textit{Submitted to proceedings of Accademia Peloritana dei Pericolanti (2016).}
\end{abstract}

\section{Introduction}
In 1884 Henri Louis Le Chatelier proposed based on a vast collection of empirical facts a general principle concerning shifts of equilibrium states of systems due to external influences, the Braun-Le Chatelier principle \cite{LeChatelier}:

\textit{
Tout système en équilibre chimique stable soumis à l’influence d’une
cause extérieure qui tend à faire varier soit sa température, soit sa
concentration (pression, concentration, nombre de molécules dans
l’unité de volume) dans sa totalité ou seulement dans quelques-unes
de ses parties, ne peut éprouver que des modifications intérieures, qui,
si elles se produisaient seules, amèneraient un changement de
température ou de condensation de signe contraire à celui résultant de
la cause extérieure.
}\footnote{English translation from \url{http://web.lemoyne.edu/giunta/lechat.html}: 
Every system in stable chemical equilibrium submitted to the influence of an exterior force which tends to cause variation, either in its temperature or its condensation (pressure, concentration, number of molecules in the unit of volume) in its totality or only in some one of its parts can undergo only those interior modifications which, if they occur alone, would produce a change of temperature, or of condensation, of a sign contrary to that resulting from the exterior force.}

From the point of view of modern equilibrium thermodynamics \cite{Gibbscw,Callen}, the principle can be seen as a consequence of concavity of entropy as function of energy, volume and number of moles, i.e. a consequence of thermodynamic stability of the equilibrium states. Indeed, nothing but thermodynamic inequalities implied by concavity of entropy is needed to derive the principle, see for instance section \S 22 of textbook \cite{Landau5}. The principle has numerous applications ranging from physical chemistry \cite{Atkins,Callen} to economics \cite{Samuelson}.

The classical Braun-Le Chatelier principle talks about shifts of equilibrium states. What if the system is in a steady non-equilibrium state, e.g. under imposed gradient of temperature? The purpose of this paper is to extend the classical principle so that shifts in stable non-equilibrium states are also covered.

To be able to discuss the non-equilibrium states, we have to choose a particular description of non-equilibrium dissipative thermodynamics. A modern approach was formulated in \cite{Grmela1997,Ottinger1997}, where non-equilibrium evolution equations are split into their reversible and irreversible parts, and the reversible part is Hamiltonian while the irreversible part is given by gradient dynamics generated by a dissipation potential $\Xi$. The gradient dynamics can be seen as an analogue of Gibbs formulation of thermodynamics, namely the dissipation potential is a function of thermodynamic forces, i.e. $\Xi(\XX)$, and thermodynamic fluxes are conjugate to the forces by Legendre transform. That setting was referred to as dissipative thermodynamics in \cite{RedExt}. To our best knowledge, such an extension of Braun-Le Chatelier principle to dissipative thermodynamics has not yet been mentioned in literature.

\section{Equilibrium Thermodynamics}\label{sec.BLC}
Consider an isolated system, state of which is given by two variables, $x$ and $y$. The former variable is supposed to evolve much faster than the latter. State of the system is also described by thermodynamic potential
\begin{equation}\label{eq.Phi}
 \Phi(x,y) = -S(x,y) + \frac{1}{T_0} E(x,y) +\frac{p_0}{T_0}V(x,y) - \frac{\mu_0}{T_0} M(x,y),
\end{equation}
$S$, $E$, $V$ and $M$ being entropy, energy, volume and mass of the system, respectively, while $T_0$, $p_0$ and $\mu_0$ stand for temperature, pressure and chemical potential, respectively, the system would attain after relaxing to thermodynamic equilibrium. 

Conjugate variables are then identified as derivatives of the thermodynamic potential, 
\begin{equation}\label{eq.forces}
 x^* = \left(\frac{\partial \Phi}{\partial x}\right)_y \mbox{ and } y^* = \left(\frac{\partial \Phi}{\partial y}\right)_y.
\end{equation}
Conjugate variables will be referred to as thermodynamic forces because as they vanish, the system approaches thermodynamic equilibrium. Indeed, taking for example energy as the state variable $x$, the conjugate variables is $x^* = -T^{-1} + T^{-1}_0$.

The system is initially in an initial state $(x_0, y_0)$, and it is then perturbed by an external force so that $y=y_0 + \delta y$. What will be the reaction of the system if only irreversible dynamics takes place? The fast variable will relax so that the fast thermodynamic force vanishes, $\XE=0$. Let us now have a look how the system behaves after the fast variables has relaxed.

Due to convexity of thermodynamic potential $\Phi$, we have the following inequalities (coming from positive definiteness of the matrix of second derivatives of $\Phi$)
\begin{subequations}\label{eq.ineq}
\begin{equation}\label{eq.Phi.convex}
\left(\frac{\partial \XE}{\partial x}\right)_y > 0, \qquad
\left(\frac{\partial \YE}{\partial y}\right)_x > 0
\end{equation}
and 
\begin{equation}\label{eq.det}
\left(\frac{\partial \XE}{\partial x}\right)_y
\left(\frac{\partial \YE}{\partial y}\right)_x
-
\left(\frac{\partial \XE}{\partial y}\right)_x
\left(\frac{\partial \YE}{\partial x}\right)_y > 0.
\end{equation}

\end{subequations}
\begin{subequations}\label{eq.Maxwell}
Moreover, from the definition of thermodynamic forces, Eq. \eqref{eq.forces}, it follows that
\begin{equation}
 \left(\frac{\partial \XE}{\partial y}\right)_{x} = 
 \left(\frac{\partial \YE}{\partial x}\right)_{y}
\end{equation}
and that
\begin{equation}\label{eq.chain}
 -1 = \left(\frac{\partial x}{\partial y}\right)_\XE \left(\frac{\partial \XE}{\partial x}\right)_y \left(\frac{\partial y}{\partial \XE}\right)_x.
\end{equation}
\end{subequations}

Let us now evaluate change in $\YE$ with respect to $y$ when keeping force $\XE$ constant. This corresponds to the state of the system where the fast variable has already relaxed so that the respective force vanishes.
\begin{eqnarray}
 \left(\frac{\partial \YE(x(\XE,y),y)}{\partial y}\right)_\XE &=& \left(\frac{\partial \YE}{\partial x}\right)_y\Big|_{(x(\XE,y),y)} \left(\frac{\partial x}{\partial y}\right)_\XE + \left(\frac{\partial \YE}{\partial y}\right)_x\Big|_{(x(\XE,y),y)}=\nonumber\\
 &\stackrel{\mbox{\eqref{eq.chain}}}{=}&\left(\frac{\partial \YE}{\partial y}\right)_x\Big|_{(x(\XE,y),y)} - \underbrace{\frac{\left(\frac{\partial \YE}{\partial x}\right)_y\left(\frac{\partial \XE}{\partial y}\right)_x}{\left(\frac{\partial \XE}{\partial x}\right)_y}\Big|_{(x(\XE,y),y)}}_{>0},
\end{eqnarray}
where the inequality follows from inequalities \eqref{eq.ineq}. Setting $\XE=0$ then leads to
\begin{equation}\label{eq.neq}
 \left(\frac{\partial \YE}{\partial y}\right)_\XE\Big|_{({\XE=0,y})} < \left(\frac{\partial \YE}{\partial y}\right)_x\Big|_{(x(\XE=0,y),y)}.
\end{equation}
This result can be interpreted as follows. Firstly, variable $y$ is perturbed. This creates force $\YE$, derivative of which is on the right hand side of inequality \eqref{eq.neq}. Due to the mixed second derivatives of $\Phi$, the fast force $\XE$ is also altered by the perturbation. The fast variable, $x$, then relaxes so that $\XE=0$ while $y$ remains constant. Derivative of $\YE$ with respect to $y$ along $\XE=0$ then becomes the left hand side of \eqref{eq.neq}, and it is lower than the original derivative along $x=const.$ The subsequent evolution of the slow variable, $y$, is then ``milder'' as the second derivative of the thermodynamic potential along $\XE=0$ is lower.

Let us now suppose that the system was initially in a state $(x_0, y_0)$ such that 
\begin{equation}
\XE(x_0,y_0) = 0.
\end{equation}
The slow variable is the perturbed as
\begin{equation}
 y = y_0 + \delta y.
\end{equation}
The force just after the perturbation can be approximated by
\begin{equation}
 \YE(x_0, y_0 + \delta y) = \YE(x_0, y_0) + \left(\frac{\partial \YE}{\partial y}\right)_x\Big|_{(x_0,y_0)}\delta y,
\end{equation}
and the force after the fast variable has relaxed by
\begin{equation}
 \YE(x(\XE=0,y_0 + \delta y), y_0 + \delta y) = \YE(x_0, y_0) + \left(\frac{\partial \YE}{\partial y}\right)_\XE\Big|_{(\XE=0,y_0)}\delta y
\end{equation}
Up to second order in $\delta y$, point $y_0$ can be replaced by $y$ in the two above formulas. Using inequality \eqref{eq.neq} then leads to 
\begin{equation}\label{eq.YY}
 |\underbrace{\YE(x_0, y_0 + \delta y) - \YE(x_0, y_0)}_{(\Delta \YE)_{x_0}}| > |\underbrace{\YE(x(\XE=0,y_0 + \delta y),y_0+\delta y) - \YE(x_0, y_0)}_{(\Delta \YE)_{\XE=0}}|,
\end{equation}
which can be interpreted as that the relaxation of the fast variable, $x$, reduces the thermodynamic force caused by the perturbation. That is the standard meaning of Braun-Le Chatelier principle, and the derivation is taken from \cite{Landau5}.

In summary, perturbation of the slow variable, $y$, leads implicitly also to perturbation of the fast variable, $x$. After the fast variable has relaxed, the thermodynamic potential has lower second derivative with respect to $y$ along $\XE=0$ than along $x=const.$ Therefore, the system experiences ``milder'' thermodynamic potential. Moreover, if the fast variable was initially relaxed, then relaxation of the variable after the perturbation leads to reducing the thermodynamic force $\YE$.

\section{Dissipative thermodynamics}
Dissipation potential is a function of thermodynamic forces $\Xi(\XX)$. The forces are no longer restricted to being conjugates of state variables, but they can be for example spatial gradients of the conjugate variables. Let us therefore consider the forces as general and yet undetermined. Thermodynamic fluxes are conjugate to the fluxes through Legendre transform
\begin{equation}
 \frac{\partial}{\partial \XX}\left(-\Xi(\XX) + \JJ\cdot\XX\right) = 0 \Rightarrow \JJ = \frac{\partial \Xi}{\partial \XX},
\end{equation}
which yields forces as functions of the fluxes, $\tilde{\XX}(\JJ)$, see e.g. \cite{RedExt}.
The Legendre transform results in a dual dissipation potential
\begin{equation}
 \Xi^*(\JJ) = -\Xi(\tilde{\XX}(\JJ)) + \JJ\cdot\tilde{\XX}(\JJ). 
\end{equation}
Backward Legendre transformations gives
\begin{equation}
 \XX = \frac{\partial \Xi^*}{\partial \JJ}.
\end{equation}
This is the setting of dissipative thermodynamics discussed for example in \cite{RedExt}. Note that analogue of classical Maxwell relations are the Maxwell-Onsager relations, generalizing the Onsager reciprocal relations to far-from-equilibrium regime. Maxwell-Onsager relations for example play the role of Eq. \eqref{eq.Maxwell} in dissipative thermodynamics.

The analysis from Sec. \ref{sec.BLC} can be extended also into the non-equilibrium thermodynamics, where the role of variables is played by forces $X$ and $Y$. Instead of conjugate variables one has fluxes $J$ and $K$ and instead of the thermodynamic potential one has the dissipation potential $\Xi$, which is a convex function of $X$ and $Y$. Forces and fluxes are related through
\begin{equation}\label{eq.fluxes}
 J = \left(\frac{\partial \Xi}{\partial X}\right)_Y \mbox{ and } K = \left(\frac{\partial \Xi}{\partial Y}\right)_X.
\end{equation}
Inequalities \eqref{eq.Phi.convex} become
\begin{subequations}
\begin{equation}\label{eq.Xi.convex}
\left(\frac{\partial J}{\partial X}\right)_Y > 0, \qquad
\left(\frac{\partial K}{\partial Y}\right)_X > 0
\end{equation}
and 
\begin{equation}\label{eq.Xi.det}
\left(\frac{\partial J}{\partial X}\right)_Y
\left(\frac{\partial K}{\partial Y}\right)_X
-
\left(\frac{\partial J}{\partial Y}\right)_X
\left(\frac{\partial K}{\partial X}\right)_Y > 0.
\end{equation}
\end{subequations}
Finally, inequality \eqref{eq.neq} becomes
\begin{equation}\label{eq.Xi.neq}
 \left(\frac{\partial K}{\partial Y}\right)_J\Big|_{({J,Y})} < \left(\frac{\partial K}{\partial Y}\right)_X\Big|_{(X(J,Y),Y)}.
\end{equation}

Suppose that the system was initially in such a state $(X_0, Y_0)$ that 
\begin{equation}
 J(X_0, Y_0) = 0.
\end{equation}
From inequality \eqref{eq.YY} it follows that
\begin{equation}\label{eq.KK}
 |\underbrace{K(X_0, Y_0 + \delta Y) - K(X_0, Y_0)}_{(\Delta K)_{X_0}}| > |\underbrace{K(X(J=0,Y_0 + \delta Y),Y_0+\delta Y) - K(X_0, Y_0)}_{(\Delta K)_{J=0}}|,
\end{equation}
which means that force $X$ relaxes so that the flux $K$ is reduced.

In summary, perturbation of the slow force, $Y$, implicitly leads to perturbation of $X$ because $J\neq 0$ due to mixed derivatives of the dissipation potential. After the fast force, $X$, has relaxed to a value given by $Y$ and $J=0$, the second derivative of the dissipation potential with respect to $Y$ along the curve $J=0$ is lower than along $X=const.$ Therefore, the system experiences ``milder'' dissipation potential after the fast force has relaxed. Moreover, if the flux $J$ is zero initially, then the relaxation of $X$ reduces the flux $K$ caused by the perturbation. That is the extended non-equilibrium thermodynamic meaning of Braun-Le Chatelier principle.

\section{Illustrations}
\subsection{Equilibrium thermodynamics}
Braun-Le Chatelier principle has a wide range of applications, which can be found in classical textbooks like \cite{Landau5,Atkins,Callen}. 
For example, let $y$ be entropy of a body and $x$ volume of the body. Entropy can be raised by heat flux into the body, and the slow thermodynamic forces is $\YE = -1/T+1/T_0$. The fast force is $\XE = \frac{p}{T} -\frac{p_0}{T_0}$. First some heat is added to the body. Volume of the body then relaxes so that $\XE=0$, and inequality \eqref{eq.YY} yields
\begin{equation}\label{eq.YY.SV}
 \left|\Delta \frac{1}{T}\right|_{V_0} > \left|\Delta\frac{1}{T}\right|_{p_0},
\end{equation}
which means that volume relaxes so that the temperature perturbation is reduced. Many similar applications including chemical composition, phase transformations or elasticity can be found in the classical textbooks \cite{Atkins, Callen, Landau5}.

\subsection{Dissipative thermodynamics}

\subsection{Transport of water and protons}
Consider now a Nafion membrane, which is a substantial part of polymer-electrolyte fuel cells, see e.g. \cite{Kjelstrup}. The membrane is subject to gradient of water chemical potential and to voltage difference, and thus there is transport of water and protons taking place inside the membrane. Those two fluxes are related to the respective thermodynamic forces through linear force-flux relations, see \cite{Kjelstrup, dGM},
\begin{subequations}
\begin{eqnarray}
\label{eq.jh}\jh &=& \Lhh \Xh + \Lhw \Xw\\
\label{eq.jw}\jw&=& \Lwh \Xh + \Lww \Xw
\end{eqnarray}
\end{subequations}
where $\Xh=-\frac{1}{T}F\nabla\Phi$ and $\Xw = -\frac{1}{T}\nabla\mu_w$. 

Moreover, Onsager reciprocity relations yield that
\begin{equation}\label{eq.LwhLhw.Onsager}
\Lhw = \Lwh,
\end{equation}
and to fulfill the second law of thermodynamics for arbitrary fluxes it is necessary that $\Lhh$ and $\Lww$ are nonnegative and that 
\begin{equation}\label{eq.posdef}
\Lhh\Lww - \Lhw\Lwh \geq 0.
\end{equation}

Expressing $\Xh$ in equation \eqref{eq.jw} by means of equation \eqref{eq.jh} and $\Xw$  in \eqref{eq.jh} by means of \eqref{eq.jw} then leads to 
\begin{subequations}
\begin{eqnarray}
\label{eq.jw2}\jw = \frac{\Lww\Lhh-\Lwh\Lhw}{\Lhh}\Xw + \frac{\Lwh}{\Lhh}\jh \\
\label{eq.jh2}\jh = \frac{\Lww\Lhh-\Lwh\Lhw}{\Lww}\Xh + \frac{\Lwh}{\Lww}\jw .
\end{eqnarray}
\end{subequations}

Comparing Eq. \eqref{eq.jh} to Eq. \eqref{eq.jh2}, we can see that
\begin{equation}\label{eq.jhjh}
 \left(\frac{\partial \jh}{\partial \Xh}\right)_{\jw} < 
 \left(\frac{\partial \jh}{\partial \Xh}\right)_{\Xw},
\end{equation}
which is a particular realization of inequality \eqref{eq.Xi.neq}. Inequality \eqref{eq.jhjh}, which was obtained explicitly here, can be interpreted so that after applying a voltage, proton and the coupled water flux arise. After water flux attains its stationary value, the change of proton flux caused by perturbation of voltage is lower than before relaxation of water flux.

Moreover, inequality \eqref{eq.KK} means that if there is initially no water flux and voltage is perturbed, the immediate proton flux is higher than the proton flux after water flux has vanished.

\subsubsection{Coupled chemical reactions}
Consider two coupled chemical reactions as for example in \cite{Lebon-Understanding}, and assume that one is much faster than the other. The system is in a steady state where the fast reaction no longer proceeds while the slow reaction proceeds due to external feed. Changing the external conditions then alters the chemical affinity of the slow reaction and (by coupling within dissipation potential) also the affinity of the fast reaction. The fast reaction is triggered. After its rate vanishes, rate of the slow reaction will be reduced. That is the meaning of inequality \eqref{eq.KK}.

Note however, that the dissipation potential might be non-convex in some region of forces, which would result in instability and phase transformation of some kind.

\subsubsection{Rheology}
Consider a Poiseuille flow through a tube. Raising the pressure gradient along the tube, the shear rate, which is a thermodynamic force, raises. However, a faster coupled dissipative process may be present in the fluid, which after relaxing to a new stationary state reduces the irreversible Cauchy stress caused by the shear rate.

When the fluid passing through the tube is isothermal and composed of two components, a natural set of state variables descring the fluid is $(\rho_1, \uu_1,\rho_2, \uu_2)$, i.e. densities and momenta of the two components. Friction between the components is expressed by Maxwell-Stefan dissipation potential
\begin{equation}
 \Xi^{(MS)} = \intd\rr \frac{1}{2}D (\uu^*_1 - \uu^*_2)^2,
\end{equation}
and friction among particles of the same constituent by the viscous dissipation potential
\begin{equation}
 \Xi^{(\nu)} = \intd\rr \sum_{i=1}^2 \frac{1}{2}\nu_i \left(\nabla\uu^*_i + (\nabla\uu^*_i)^T\right)^2.
\end{equation}
The complete dissipation potential is then sum of the two above parts, $\Xi = \Xi^{(MS)} + \Xi^{(\nu)}$.

Changing variables to 
\begin{subequations}
\begin{eqnarray}
\rho&=&\rho_1+\rho_2, \\
c&=&\frac{\rho_1}{\rho_1+\rho_2},\\
\uu &=& \uu_1 + \uu_2,\\
\ww &=& \frac{\rho_2}{\rho_1+\rho_2}\uu_1 - \frac{\rho_1}{\rho_1+\rho_2}\uu_2
\end{eqnarray}
as in \cite{Elafif1999}, conjugate momenta transform as
\begin{subequations}
\begin{eqnarray} 
\uu^* &=& \left(\frac{\delta \Phi}{\delta \uu}\right)_{\ww} 
= c \uu^*_1 +(1-c)\uu^*_2 \\
\ww^* &=& \left(\frac{\delta \Phi}{\delta \ww}\right)_{\uu} 
= \uu^*_1 - \uu^*_2.
\end{eqnarray}
\end{subequations}
\end{subequations}
The total dissipation potential then becomes
\begin{multline}
 \Xi = \intd\rr 
 \frac{1}{2}D (\ww^*)^2\\
 +\frac{1}{2} \nu_1\left(\nabla (\uu^* + (1-c)\ww^*) + \nabla (\uu^* + (1-c)\ww^*)^T\right)^2 \\
 +\frac{1}{2} \nu_2\left(\nabla (\uu^* - c\ww^*) + \nabla (\uu^* - c\ww^*)^T\right)^2,
\end{multline}
where coupling between shear rate $\nabla\uu^*$ and diffusion $\ww^*$, $\nabla \ww^*$ is evident. 

When diffusion relaxes faster than overall momentum, as for example in aerodynamics, where air is often considered as a one-component fluid described by Navier-Stokes-Fourier equations, the Braun-Le Chatelier principle in dissipative thermodynamics applies. 

It was experimentally observed, see e.g. \cite{Weeks}, that colloidal particles dispersed in a fluid tend to accumulate in the center of the tube. The phenomenon is called particle migration. If the resulting irreversible stress is reduced by the migration, the above discussed coupling between overall shear rate and diffusion could be responsible for the migration via the Braun-Le Chatelier principle.

\section{Conclusion}
Braun-Le Chatelier principle is a direct consequence of convexity of entropy (or the corresponding thermodynamic potential), and it expresses how equilibrium states shift when external influence affects the system. 
Convexity of dissipation potential then leads to analogical results, showing how steady non-equilibrium states shift when thermodynamic forces are altered. More precisely, once the slow thermodynamic force is perturbed, for example by varying boundary conditions, the fast thermodynamic force is also perturbed by coupling through the dissipation potential. The fast force then relaxes to such a value that the conjugate thermodynamic flux vanishes, and the initial perturbation of the slow force is reduced by relaxation of the fast force. That is the extension of Braun-Le Chatelier principle to dissipative thermodynamics. The principle is then demonstrated on ion-conducting membranes, chemical kinetics and rheology of suspensions.

\section*{Acknowledgement}
This project was supported by Natural Sciences and Engineering Research Council of Canada (NSERC).


\end{document}